\documentclass{article}
\usepackage{slashed,epsfig,amsmath,amssymb,enumitem}
\usepackage[papersize={8.5in,11in}]{geometry}
\usepackage{graphicx}
\begin{document}
\title{Black Holes in Modified Gravity (MOG)}
\author{J. W. Moffat\\~\\
Perimeter Institute for Theoretical Physics, Waterloo, Ontario N2L 2Y5, Canada\\
and\\
Department of Physics and Astronomy, University of Waterloo, Waterloo,\\
Ontario N2L 3G1, Canada}
\maketitle




\begin{abstract}
The field equations for Scalar-Tensor-Vector-Gravity (STVG) or modified gravity (MOG) have a static, spherically symmetric black hole solution determined by the mass $M$ with two horizons. The strength of the gravitational constant is $G=G_N(1+\alpha)$ where $\alpha$ is a parameter. A regular singularity-free MOG solution is derived using a nonlinear field dynamics for the repulsive gravitational field component and a reasonable physical energy-momentum tensor. The Kruskal-Szekeres completion of the MOG black hole solution is obtained. The Kerr-MOG black hole solution is determined by the mass $M$, the parameter $\alpha$ and the spin angular momentum $J=Ma$. The equations of motion and the stability condition of a test particle orbiting the MOG black hole are derived, and the radius of the black hole photosphere and the shadows cast by the Schwarzschild-MOG and Kerr-MOG black holes are calculated. A traversable wormhole solution is constructed with a throat stabilized by the repulsive component of the gravitational field.
\end{abstract}

\maketitle

\section{Introduction}

The Scalar-Tensor-Vector (STVG) modified gravitational (MOG) theory~\cite{Moffat1} has successfully explained the solar system observations~\cite{Moffat2}, the rotation curves of galaxies~\cite{MoffatRahvar1,TothMoffat} and the dynamics of galactic clusters~\cite{MoffatRahvar2,BrownsteinMoffat}, as well as describing the growth of structure, the matter power spectrum and the cosmic microwave background (CMB) acoustical power spectrum data~\cite{Moffat3}. In the following, we investigate the nature of a black hole solution to the field equations. The static spherically symmetric solution describes the final stage of the collapse of a body in terms of an enhanced gravitational constant $G$ and a gravitational repulsive force with a charge $Q=\sqrt{\alpha G_N}M$, where $\alpha$ is a parameter defined by $G=G_N(1+\alpha)$, $G_N$ is Newton's constant and $M$ is the total mass of the black hole. The black hole has two horizons for $\alpha > 0$ and there is no horizon-free solution with a naked singularity. The vector field $\phi_\mu$ produces a repulsive gravitational force in the theory, which repels a test particle falling through the inner horizon preventing it from reaching the singularity at $r=0$.

A MOG static spherically symmetric solution regular at $r=0$ is obtained, which has two horizons when $\alpha < \alpha_{\rm crit}=0.673$ and the metric is an effective de Sitter (anti-de Sitter) spacetime as $r\rightarrow 0$. For $\alpha > \alpha_{\rm crit}$ the spacetime has no horizons and is regular at $r=0$. The collapsed object described by this spacetime has the topological structure $S^3$ compared with the singular MOG solution which has the topology $S^2\times R$. The regular solution describes a grey hole.

A derivation of the Kruskal-Szekeres~\cite{Kruskal,Szekeres} analytic completion of the black hole solutions is obtained. The photosphere produced by photons orbiting around the black hole is calculated in terms of the charge $Q=\sqrt{\alpha G_N}M$. The Kerr-MOG black hole solution is determined by the spin angular momentum parameter $a$, the mass $M$ and the parameter $\alpha$. The motion and stability condition of a test particle orbiting the black hole are derived.

A traversable MOG wormhole solution is obtained with a wormhole throat stabilized by the balance of attractive and repulsive gravitational forces. The wormhole allows observers to causally connect distant parts of asymptotically flat regions of the universe or connect to different universes.

\section{Field Equations and MOG Black Hole Solution}

In the following, we will be investigate the STVG field equation for the metric tensor $g_{\mu\nu}$~\cite{Moffat1}:
\begin{equation}
\label{phiFieldEq}
R_{\mu\nu}=-8\pi GT_{\phi\mu\nu},
\end{equation}
where we have set $c=1$ and assumed that the measure of gravitational coupling $G=G_N(1+\alpha)$ is constant, $\partial_\nu G = 0$, and that the matter energy-momentum tensor $T_{M\mu\nu}=0$. We have $B_{\mu\nu}=\partial_\mu\phi_\nu-\partial_\nu\phi_\mu$ and the energy-momentum tensor for the $\phi_\mu$ vector field is given by \footnote{We have assumed that the potential $V(\phi)$ in the definition in~\cite{Moffat1} of the energy-momentum tensor $T_{\phi\mu\nu}$ is zero.}
\begin{equation}
\label{Tphi}
T_{\phi\mu\nu}=-\frac{1}{4\pi}({B_\mu}^\alpha B_{\nu\alpha}-\frac{1}{4}g_{\mu\nu}B^{\alpha\beta}B_{\alpha\beta}).
\end{equation}
The dimensionless constant $\omega$ included in the STVG action in~\cite{Moffat1} as a scalar field is now treated as a constant and in the following $\omega=1$. When the parameter $\alpha$ in the definition of the gravitational source charge $Q=\sqrt{\alpha G_N}M$ of the vector field $\phi_\mu$ vanishes, then STVG reduces to GR. We neglect the mass of the $\phi_\mu$ field, for in the determination of galaxy rotation curves and galactic cluster dynamics $\mu=0.042\,(\rm kpc)^{-1}$, which corresponds to the vector field $\phi_\mu$ mass $m_\phi=2.6\times 10^{-28}$ eV~\cite{MoffatRahvar1,TothMoffat,MoffatRahvar2}. The smallness of the $\phi_\mu$ field mass in the present universe justifies our ignoring it when solving the field equations for compact objects such as neutron stars and black holes. However, the scale $\mu=0.042\,{\rm kpc}^{-1}$ does play an important role in fitting galaxy rotation curves and cluster dynamics. We also need the vacuum field equations:
\begin{equation}
\label{Bequation}
\nabla_\nu B^{\mu\nu}=\frac{1}{\sqrt{-g}}\partial_\nu(\sqrt{-g}B^{\mu\nu})=0,
\end{equation}
and
\begin{equation}
\label{Bcurleq}
\nabla_\sigma B_{\mu\nu}+\nabla_\mu B_{\nu\sigma}+\nabla_\nu B_{\sigma\mu}=0,
\end{equation}
where $\nabla_\nu$ is the covariant derivative with respect to the metric tensor $g_{\mu\nu}$.

We adopt the static spherically symmetric metric:
\begin{equation}
ds^2=\exp(\nu)dt^2-\exp(\lambda)dr^2-r^2d\Omega^2,
\end{equation}
where $d\Omega^2=d\theta^2+\sin^2\theta d\phi^2$.
We have for the static solution $\phi_0\neq 0$ and $\phi_1=\phi_2=\phi_3=0$,
and from (\ref{Bequation}) we get
\begin{equation}
\partial_r(\sqrt{-g}B^{0r})=-\sin\theta\partial_r(\exp[-(\lambda+\nu)/2]r^2\phi_0')=0,
\end{equation}
where $\phi_0'=\partial_r\phi_0$. Integrating this equation, we get
\begin{equation}
\phi_0'=\exp[(\lambda+\nu)/2]\frac{Q}{r^2},
\end{equation}
where $Q$ is the gravitational source charge of the $B_{\mu\nu}$ field. Moreover, we have
\begin{equation}
{T_{\phi1}}^1=-{T_{\phi2}}^2=-{T_{\phi3}}^3={T_{\phi0}}^0=\frac{1}{2}\exp(-\lambda-\nu)(\phi_0')^2=\frac{1}{8\pi}\frac{Q^2}{r^4}.
\end{equation}
Setting $\gamma=\exp(\nu)$ and solving the field equations (\ref{phiFieldEq}), we obtain $\lambda'=-\nu'$ and
\begin{equation}
\gamma+r\gamma'=1-\frac{GQ^2}{r^2},
\end{equation}
\begin{equation}
r\gamma=r+\frac{GQ^2}{r}-2GM,
\end{equation}
where $2GM$ is a constant of integration. The gravitational field metric is given by
\begin{equation}
\label{metric}
ds^2=\biggl(1-\frac{2GM}{r}+\frac{GQ^2}{r^2}\biggr)dt^2-\biggl(1-\frac{2GM}{r}
+\frac{GQ^2}{r^2}\biggr)^{-1}dr^2-r^2d\Omega^2,
\end{equation}
where $G=G_N(1+\alpha)$. This has the form of the static, point particle Reissner-Nordstr\"om solution for an electrically charged body~\cite{Reissner,Nordstrom}, but now the charge $Q > 0$ is of gravitational origin \footnote{Astrophysical bodies including black holes are electrically neutral.}.

We postulate that the charge $Q$ is proportional to the mass of the source particle, $Q=\kappa M$. We have
\begin{equation}
\kappa=\pm\sqrt{\alpha G_N},
\end{equation}
where we define $\alpha=(G-G_N)/G_N$~\cite{MoffatRahvar1}. This yields the physical value for the charge $Q$:
\begin{equation}
\label{MOGcharge}
Q=\pm\sqrt{\alpha G_N}M.
\end{equation}
To maintain a repulsive, gravitational Yukawa force when the mass parameter $\mu$ is non-zero, necessary to describe physically stable stars, galaxies, galaxy clusters and agreement with solar system observational data, we choose the positive value for the square root in (\ref{MOGcharge}) giving $Q > 0$. We now obtain for the metric (\ref{metric}):
\begin{equation}
\label{MOGmetric}
ds^2=\biggl(1-\frac{2GM}{r}+\frac{\alpha G_NGM^2}{r^2}\biggr)dt^2-\biggl(1-\frac{2GM}{r}+\frac{\alpha G_NGM^2}{r^2}\biggr)^{-1}dr^2-r^2d\Omega^2,
\end{equation}
where as before $G=G_N(1+\alpha)$. The metric reduces to the Schwarzschild solution when $\alpha=0$. To determine the solar system tests predicted by the metric, we must use the weak field and slow velocity test particle equation of motion in which the repulsive Yukawa gravitational force cancels the attractive force at the scale of the solar system and results in agreement of the theory with solar system tests (see Eq.(\ref{eqmotion}) in Section 4 and~\cite{Moffat2}).

As for the Reissner-Nordstr\"om and the Kerr solution~\cite{Kerr} the MOG black hole has two horizons:
\begin{equation}
\label{2horizons}
r_{\pm}=G_NM\biggl[1+\alpha\pm \biggl(1+\alpha\biggr)^{1/2}\biggr].
\end{equation}
We obtain for $\alpha=0$ the Schwarzschild radius event horizon $r_+=r_s=2G_NM$.
The inner horizon $r_{-}$ is a Cauchy horizon and is expected to be unstable as in the case of the Cauchy horizon for the Reissner-Nordstr\"om metric~\cite{Matzner}. The singularity at $r=0$ in the black hole can be removed by quantum gravity~\cite{Moffat4,Moffat5,Moffat6}.

We can also consider singularity-free classical solutions of our modified field equations. A black hole solution regular at $r=0$ has been derived by Ay\'on-Beato and Garc\'{i}a~\cite{Garcia1,Garcia2,Garcia3}. It is an exact solution of the Einstein field equations with mass and electric charge as two parameters in the solution, in which the source is a nonlinear electrodynamic field satisfying the weak energy condition $T_{\mu\nu}V^\mu V^\nu > 0$, where $V^\mu$ is a timelike vector. The metric has the form of a Bardeen-type black hole~\cite{Bardeen1,Borde1,Borde2,Ansoldi} with a physical source and energy-momentum tensor regular at $r=0$.

A metric solution regular at $r=0$ is obtained from the MOG field equations for a reasonably physical energy-momentum tensor $T_{\phi\mu\nu}$ and a nonlinear $B_{\mu\nu}$ field dynamics~\cite{Garcia1}:
\begin{eqnarray}
\label{BardeenMOGmetric}
ds^2&=&\biggl(1-\frac{2GMr^2}{(r^2+\alpha G_NGM^2)^{3/2}}+\frac{\alpha G_NGM^2r^2}{(r^2+\alpha G_NGM^2)^2}\biggr)dt^2\nonumber\\
&-&\biggl(1-\frac{2GMr^2}{(r^2+\alpha G_NGM^2)^{3/2}}+\frac{\alpha G_NGM^2r^2}{(r^2+\alpha G_NGM^2)^2}\biggr)^{-1}dr^2-r^2d\Omega^2.
\end{eqnarray}
The associated gravitational $B_{0r}$ field is given by
\begin{equation}
\label{GarciaB}
B_{0r}=(\sqrt{\alpha G_N}M)r^4\biggl(\frac{r^2-5\alpha G_NGM^2}{(r^2+\alpha G_NGM^2)^4}+\frac{15}{2}\frac{GM}{(r^2+\alpha G_NGM^2)^{7/2}}\biggr).
\end{equation}
The solution (\ref{BardeenMOGmetric}) behaves asymptotically like the solution (\ref{MOGmetric}):
\begin{equation}
g_{00}=1-\frac{2GM}{r}+\frac{\alpha G_NGM^2}{r^2}+{\cal O}(1/r^3),
\end{equation}
and
\begin{equation}
B_{0r}=\frac{\sqrt{\alpha G_N}M}{r^2}+{\cal O}(1/r^3).
\end{equation}
The $B_{0r}$ field (\ref{GarciaB}) is regular at $r=0$. The nonlinear $B_{\mu\nu}$ field dynamics behaves for increasing $r$ like the MOG gravitational weak field dynamics and the spacetime geometry is determined by the parameter $\alpha$ and the total mass $M$. The metric solution has finite values for the curvature scalar $R={R_\mu}^\mu$ and the Kretchmann invariant $R_{\mu\nu\lambda\sigma}R^{\mu\nu\lambda\sigma}$ at $r=0$. It is asymptotically flat as $r\rightarrow\infty$ and is the Schwarzschild metric for large $r$. For small $r$ the metric behaves as
\begin{equation}
\label{deSitterMog}
ds^2=(1-\frac{1}{3}\Lambda r^2)dt^2-(1-\frac{1}{3}\Lambda r^2)^{-1}dr^2-r^2d\Omega^2,
\end{equation}
where the effective cosmological constant $\Lambda$ is given by
\begin{equation}
\Lambda=\frac{3}{G_N^2M^2}\biggl(\frac{\alpha^{1/2}-2}{\alpha^{3/2}(1+\alpha)}\biggr).
\end{equation}
This means that the interior material of the regular MOG black hole satisfies the vacuum equation of state $p=-\rho$, where $p$ and $\rho=\rho_{\rm vac}$ denote the pressure and the vacuum density, respectively. The effective cosmological constant $\Lambda$ can be positive or negative depending on the magnitude of $\alpha$, so that the interior of the black hole is described by either a de Sitter or anti-de Sitter spacetime.

The horizons determined by $g_{00}=0$ for the regular MOG black hole are given by the real solutions of the equation:
\begin{equation}
r^8+2(3\alpha G_NGM^2-2G^2M^2)r^6+\alpha G_NGM^2(11\alpha G_NGM^2-4G^2M^2)r^4+6(\alpha G_NGM^2)^3r^2+(\alpha G_NGM^2)^4=0.
\end{equation}
It is sufficient to consider the positive roots of this equation and it has two or no positive zeros. For $\alpha < \alpha_{\rm crit}=0.673$ there are two horizons, an outer horizon $r_+$ and an inner Cauchy horizon $r_-$. For $\alpha > \alpha_{\rm crit}$ there are {\it no horizons and there is no naked singularity at $r=0$}. The topology of the spacelike sections for the MOG spacetime described by the metric (\ref{MOGmetric}) is $S^3\times R$ throughout. In the regular MOG spacetime described by the metric (\ref{BardeenMOGmetric}) the topology switches from $S^3\times R$ to $S^3$~\cite{Borde1,Borde2}.

\section{The Kruskal-Szekeres Coordinates}

\begin{figure}
\centering \includegraphics[scale=0.75]{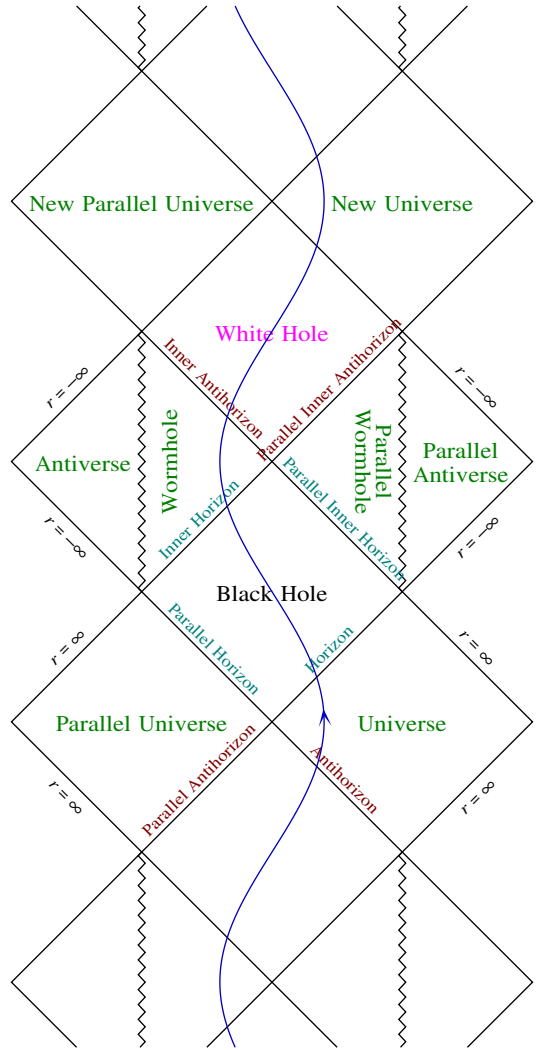}
\caption{\label{fig:penrose}The Penrose diagram for the singular MOG black hole solution (image by Andrew Hamilton).}
\end{figure}

We write the metric in the form:
\begin{equation}
\label{standardmetric}
ds^2=\gamma dt^2-\gamma^{-1}dr^2-r^2d\Omega^2,
\end{equation}
where
\begin{equation}
\gamma=1-\frac{2GM}{r}+\frac{\alpha G_NGM^2}{r^2}.
\end{equation}
We determine a simultaneous transformation of $r$ and $t$ to new coordinates $u(r,t)$ and $v(r,t)$ for which the light cones are lines with slope $\pm 1$~\cite{Kruskal,Szekeres,Brill,MTW,Hawking}. In such coordinates the metric (\ref{standardmetric}) takes the form:
\begin{equation}
ds^2=f^2(u,v)(dv^2-du^2)-r^2(u,v)d\Omega^2.
\end{equation}
The solution for the $u$ and $v$ coordinates in which the metric is nonsingular at the horizon yields coordinate patches, corresponding to the two real roots of (\ref{2horizons}). The Penrose diagram showing the singular MOG black hole geometry is displayed in Fig.~\ref{fig:penrose}. The metric written in isotropic coordinates has the form:
\begin{equation}
ds^2=\biggl[\biggl(1+\frac{2GM}{2\rho}\biggr)^2-\frac{\alpha G_NGM^2}{(2\rho)^2}\biggr]^2(d\rho^2+\rho^2d\Omega^2),
\end{equation}
where
\begin{equation}
r=\rho\biggl[\biggl(1+\frac{2GM}{2\rho}\biggr)^2-\frac{\alpha G_NGM^2}{(2\rho)^2}\biggr].
\end{equation}

The metric can be imbedded in 4-dimensional space, giving a surface of revolution obtained by rotating a parabola about a line perpendicular to its axis. The curve to be rotated in the $\theta$ and $\phi$  directions is given, in the ${\bar r}-r$ plane by
\begin{equation}
{\bar r}(r)=\int dr\biggl[\frac{(2GMr-\alpha G_NGM^2)}{(r-r_1)(r-r_2)}\biggr]^{1/2},
\end{equation}
where $r_1$ and $r_2$ denote the minimum values of $r$ in two coordinate patches of the MOG Penrose diagram.

The solution for $f^2$ takes the form:
\begin{equation}
f^2(r)=\biggl(1-\frac{2GM}{r}+\frac{\alpha G_NGM^2}{r^2}\biggr)\frac{\exp(-2br^*)}{4K^2b^2},
\end{equation}
where $dr^*=\gamma^{-1}(r)dr$, $b$ is a constant, $K$ is an arbitrary scale factor and $f^2(r)$ is regular and positive throughout the coordinate patch.

The maximal analytic extension of the regular MOG black hole solution with the metric (\ref{BardeenMOGmetric}) and $\alpha < \alpha_{\rm crit}$ can be derived using the form of the standard metric (\ref{standardmetric}) and the same method used to obtain the Kruskal-Szekeres extension of the singular MOG solution.\\

\section{Test Particle Motion}
Consider the motion of test particles of mass $m$ and charge $q=\kappa m=\sqrt{\alpha G_N}m$. The equations of motion are given by
\begin{equation}
\label{eqmotion}
\frac{d^2x^\mu}{ds^2}+{\Gamma^\mu}_{\alpha\beta}\frac{dx^\alpha}{ds}\frac{dx^\beta}{ds}=\kappa{B_\sigma}^\mu\frac{dx^\sigma}{ds}.
\end{equation}
Because the test particle mass $m$ has canceled, the motion of a test particle satisfies the weak equivalence principle. Photons with mass $0$ travel along null geodesics with the equation of motion
\begin{equation}
\label{photoneqmotion}
\frac{d^2x^\mu}{ds^2}+{\Gamma^\mu}_{\alpha\beta}\frac{dx^\alpha}{d\lambda}\frac{dx^\beta}{d\lambda}=0,
\end{equation}
where $\lambda$ is an affine parameter.

From the equations of motion of a test particle (\ref{eqmotion}), we get after integration in polar coordinates with $\theta=\pi/2$:
\begin{equation}
J=r^2\frac{d\phi}{ds},\quad \frac{dt}{ds}=\frac{C}{\gamma},
\end{equation}
where $J$ and $C$ are constants of integration and $J$ is the angular momentum. Differentiating the metric (\ref{standardmetric}) with respect to $s$, we obtain
\begin{equation}
\gamma\biggl(\frac{dt}{ds}\biggr)^2-\gamma^{-1}\biggl(\frac{dr}{ds}\biggr)^2-r^2\biggl(\frac{d\phi}{ds}\biggr)^2=1,
\end{equation}
which has the form of an energy integral. We can write this equation as
\begin{equation}
\label{MOGenergyeq}
\biggl(\frac{dr}{ds}\biggr)^2+\frac{r^2}{1-2GM/r+\alpha G_NGM^2/r^2}=\frac{r^4(C^2-1)}{J^2}.
\end{equation}
The term $\alpha G^2M^2/r^2$ dominates as $r\rightarrow 0$ and we see that the test particle can never reach $r=0$.
In contrast to the analytic continuation of the Schwarzschild metric for a point mass, the throat of a wormhole describing the particle pulsates in time and does not pinch off. No test particle ever hits the singularity at $r=0$. The result that the modified gravity metric is described by a pulsating throat, is in accord with the dynamic balance between the gravitational attraction and the repulsive gravity pressure (anti-gravity) produced by the $B_{\mu\nu}$ field.

When an observer falls freely through the MOG black hole horizon and then subsequently falls through the inner Cauchy horizon, the repulsive gravity pressure propels the observer into a white hole in another universe as described in Fig. 1.

\section{Kerr-MOG Black Hole}

In addition to mass $M$, we shall include in our solution the spin angular momentum $J=Ma$:
\begin{equation}
ds^2=\frac{\Delta}{\rho^2}(dt-a\sin^2\theta d\phi)^2-\frac{\sin^2\theta}{\rho^2}[(r^2+a^2)d\phi-adt]^2-\frac{\rho^2}{\Delta}dr^2-\rho^2d\theta^2,
\end{equation}
where
\begin{equation}
\Delta=r^2-2GMr+a^2+\alpha G_NGM^2,\quad \rho^2=r^2+a^2\cos^2\theta.
\end{equation}
The spacetime geometry is axially symmetric around the $z$ axis~\cite{Newman}. Horizons are determined by the roots of $\Delta=0$:
\begin{equation}
r_\pm=G_N(1+\alpha)M\biggl[1\pm\sqrt{1-\frac{a^2}{G_N^2(1+\alpha)^2M^2}-\frac{\alpha}{1+\alpha}}\biggr].
\end{equation}
An ergosphere horizon is determined by $g_{00}=0$:
\begin{equation}
r_E=G_N(1+\alpha)M\biggl[1+\sqrt{1-\frac{a^2\cos^2\theta}{G_N^2(1+\alpha)^2M^2}-\frac{\alpha}{1+\alpha}}\biggr].
\end{equation}

Our Kerr-MOG solution of the MOG field equations is fully characterized by the mass $M$, spin parameter $a$ and the parameter $\alpha$. According to the no-hair theorem~\cite{Israel,Carter,Hawking1,Robinson}, no closed time-like curves exist outside of the horizon. Multipole moments of the Kerr-MOG spacetime $M_l$ and $J_l$ exist. The mass multipole moments $M_l$ vanish if $l$ is odd, and the set of current multipole moments $S_l$ vanish if $l$ is even. The no-hair theorem can then be stated for $\alpha > 0$ as~\cite{Geroch,Hansen}:
\begin{equation}
M_l+iS_l=M(ia)^l.
\end{equation}

The radial motion of test particles with mass $m$ in the equatorial plane $\theta=\pi/2$ is governed by~\cite{MTW}:
\begin{equation}
A(r)E^2-2B(r)E+D(r)-r^4(p^r)^2=0,\quad E=\frac{B(r)+\sqrt{B(r)^2-A(r)D(r)+A(r)r^4(p^r)^2}}{A(r)},
\end{equation}
where $E$ is the energy, $p^r$ is the radial momentum $p^r=mdr/ds$ and $A,B,D$ are given by
\begin{equation}
A(r)=(r^2+a^2)^2-a^2\Delta,
\end{equation}
\begin{equation}
B(r)=(Ja+\alpha G_NmMr)(r^2+a^2)-Ja\Delta,
\end{equation}
\begin{equation}
D(r)=(Ja+\alpha G_NmMr)^2-\Delta J^2-m^2r^2\Delta.
\end{equation}

The effective potential $V(r)$ is the minimum allowed value of the energy $E$ at radius $r$:
\begin{equation}
V(r)=\frac{B(r)+\sqrt{B^2(r)-A(r)D(r)}}{A(r)}.
\end{equation}
The permitted regions for a particle of energy $E$ are the regions with $V\leq E$, and the turning points for $p^r=mdr/ds=0$ occur when $V(r)=E$. Stable circular test particle orbits occur at the minima of $V(r)$. The irreducible mass of the Kerr-MOG black hole is given by
\begin{equation}
M_{\rm ir}=\frac{1}{2}[(G_N(1+\alpha)M+\sqrt{G_N^2(1+\alpha)^2M^2-\alpha G_N^2(1+\alpha)M^2-a^2)^2}+a^2]^{1/2}.
\end{equation}
We have $\delta M_{\rm ir} > 0$, so that no injection of small lumps of matter can ever reduce the irreducible mass of a Kerr-MOG black hole. This is consistent with the second law of black hole dynamics, for the surface area of the black hole is ${\cal A}=16\pi G_N^2(1+\alpha)^2M_{\rm ir}^2$.

The radius of the photosphere determined by our metric (\ref{MOGmetric}) is given by
\begin{equation}
r_{ps}=\frac{3}{2}G_N(1+\alpha)M\biggl(1+\sqrt{1-\frac{8\alpha}{9(1+\alpha)}}\biggr).
\end{equation}
It is necessary to take into account the screening mechanism for photon trajectories in MOG~\cite{Moffat2}. However, this screening mechanism holds for weak gravitational fields and may not hold for the strong gravitational fields near black holes. When the impact parameter $r_0$ takes values near $r_{ps}$ the angle of deflection $\delta\theta_{\rm def} > 2\pi$, whereby the light ray can turn around the black hole several times before reaching the observer. We have two infinite sets of images, one produced by clockwise winding around the black hole for $\delta\alpha > 0$, and the other one by counter-clockwise winding for $\delta\theta_{\rm def} < 0$. These images are located at the same side and at the opposite side of the source.

\section{Blackhole Shadow Image and Lensing}

A remarkable prediction of strong gravity is the existence of a region outside a black hole horizon (or no horizon), in which there are no closed photon orbits. The photons that cross this region are removed from the observable universe. This phenomenon results in a shadow (silhouette) imprinted by a black hole on the bright emission that exists in its vicinity and the apparent shape of a black hole is defined by the boundary of the black hole~\cite{Bardeen2,Luminet,deVries,Fish,Lu,Johannsen,Psaltis}. The Hamilton-Jacobi equation determines the geodesics for a given geometry~\cite{deVries,Hioki,Amarilla,Lammerzahl}:
\begin{equation}
\frac{\partial S}{\partial\lambda}=-\frac{1}{2}g^{\mu\nu}\frac{\partial S}{\partial x^\mu}\frac{\partial S}{\partial x^\nu},
\end{equation}
where $\lambda$ is an affine parameter and $S$ is the Jacobi action. The action $S$ can be written as
\begin{equation}
S=\frac{1}{2}m^2\lambda-Et+J_z\phi+S_r(r)+S_\theta(\theta),
\end{equation}
where $m$ the test particle mass, $E$ is the energy, $J_z$ is the angular momentum in the direction of the axis of symmetry. For a photon null geodesic, we have $m=0$ and from the Hamilton-Jacobi equations of motion, we obtain
\begin{equation}
\Sigma\frac{dt}{d\lambda}=a(J_z-aE\sin^2\theta)+\frac{r^2+a^2}{\Delta}[(r^2+a^2)E-aJ_z],
\end{equation}
\begin{equation}
\Sigma\frac{d\phi}{d\lambda}=\frac{J_z}{\sin^2\theta}-aE+\frac{a}{\Delta}[(R^2+a^2)E-aJ_z],
\end{equation}
\begin{equation}
\label{Requation}
\Sigma\frac{dr}{d\lambda}=\sqrt{{\cal R}},
\end{equation}
\begin{equation}
\Sigma\frac{d\theta}{d\lambda}=\sqrt{\Theta}.
\end{equation}
Here, $\Sigma=r^2+a^2\cos^2\theta$ and ${\cal R}$ and $\Theta$ are given by
\begin{equation}
{\cal R}=[(r^2+a^2)E-aJ_z]^2-\Delta[{\cal K}+(J_z-aE)^2],\quad \Theta={\cal K}+\cos^2\theta\biggl(a^2E^2-\frac{J_z^2}{\sin^2\theta}\biggr),
\end{equation}
where ${\cal K}$ is a constant. The spacetime is asymptotically flat, so that photon paths are straight lines at infinity. The photon trajectories are determined in terms of two impact parameters $\xi$ and $\eta$, which are defined by $\xi=J_z/E$ and $\eta={\cal K}/E^2$. The photon orbits with constant $r$ are derived from (\ref{Requation}) and yield the boundary of the black hole shadow. The circular photon orbits are determined by the conditions ${\cal R}(r)=0$ and $d{\cal R}/dr=0$ and we get
\begin{equation}
\xi(r)=\frac{G_N(1+\alpha)M(r^2-a^2)-r\alpha G_N^2(1+\alpha)M^2-r\Delta}{a(r-G_N(1+\alpha)M)},
\end{equation}
\begin{equation}
\eta(r)=\frac{r^2}{a^2(r-G_N(1+\alpha)M)^2}[4G_N(1+\alpha)Mr(a^2+\alpha G_N^2(1+\alpha)M^2)-4\alpha G_N^2(1+\alpha)M^2\Delta
$$ $$
-r^2(r-3G_N(1+\alpha)M)^2].
\end{equation}
These equations determine parametrically the critical locus $(\xi,\eta)$ and give the constants of the motion for the photon orbits of constant radius.

\begin{figure}
\centering \includegraphics[scale=0.4]{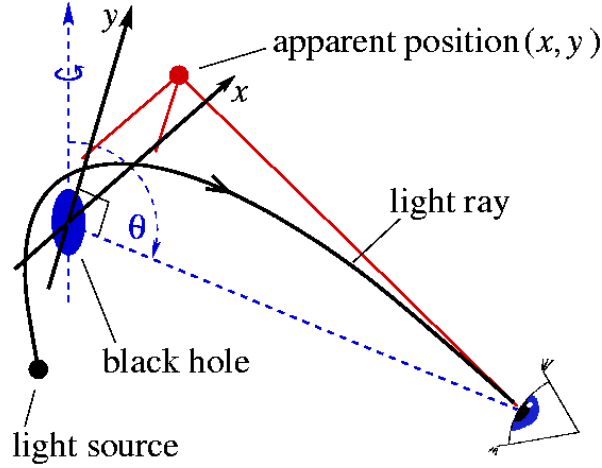}\\
\caption{\label{fig:shadow}The apparent position of a light ray with respect to the observer's projection plane in the x,y coordinates containing the center of the spacetime: x denotes the apparent distance from the rotation axis, and y the projection of the rotation axis itself (dashed line). The angle $\theta$ denotes the angle of latitude, reaching from the north pole at $\theta = 0$ to the south pole at $\theta = \pi$ (image by Math@IT).}
\end{figure}

We use celestial coordinates $x$ and $y$ to determine the black hole shadow. We consider an observer far away from the black hole and $\theta$ is the angular coordinate of the observer corresponding to the angle between the axis of rotation of the black hole and the line of sight of the observer as described in Fig.~\ref{fig:shadow}.
The celestial coordinates $x$ and $y$ of the shadow image are given by
\begin{equation}
x=-\xi\csc\theta,\quad y=\pm\sqrt{\eta+a^2\cos^2\theta-\xi^2\cot^2\theta}.
\end{equation}

The photon orbit is parameterized by the conserved quantities $(\xi,\eta)$ with the observer's polar angle $\theta$. Letting $\xi$ and $\eta$ take all possible values with a fixed $\theta$, the photon capture region is determined by the impact parameter space $(b_x,b_y)$ for the given $\theta$. The photon capture region can be regarded as the apparent shadow shape of the black hole. The shape of the black hole shadow is obtained by plotting $y$ versus $x$.\\

The lensing equation can be written as~\cite{Ellis}:
\begin{equation}
\tan\beta=\tan\theta-\frac{D_{ls}}{D_{os}}(\tan\theta+\tan(\delta\alpha-\theta)),
\end{equation}
where $\beta$ and $\theta$ are the angular source and image positions, respectively, $\delta\alpha$ is the deflection angle and $D_{ls}$ and $D_{os}$ are the distances from the lens to the source and from the observer to the source, respectively. The deflection angle is~\cite{Eiroa}:
\begin{equation}
\delta\alpha(r_0)=2\int^\infty_{r_0}drr\biggl[\biggl(r/r_0\biggr)^2\biggl(1-2G_N(1+\alpha)M/r_0+\alpha G_N^2(1+\alpha)M^2/r_0^2\biggr)
$$ $$
-\biggl(1-2G_N(1+\alpha)M/r+\alpha G_N^2(1+\alpha)M^2/r^2\biggr)\biggr]^{1/2}-\pi,
\end{equation}
where $r_0$ is the distance of closest approach. The impact parameter is given by
\begin{equation}
J(r_0)=\frac{r_0}{\sqrt{1-2G_N(1+\alpha)M/r_0+\alpha G_N^2(1+\alpha)M^2/r_0^2}},
\end{equation}
where $J(r_0)=D_{ol}\sin\theta$ with $D_{ol}$ the distance from the observer to the lens.

\section{Traversable MOG Wormhole}

The idea of a worm hole in spacetime was realized in 1916 by Flamm~\cite{Flamm} who recognized that the hypersurfaces of constant Killing time through a Schwarzschild spacetime, embedded in an Euclidean space, demonstrated that these hypersurfaces consist of two asymptotically flat sheets. These sheets are connected by an Einstein-Rosen bridge~\cite{EinsteinRosen}, or throat, called a ``wormhole'' by Wheeler~\cite{Wheeler}. The possibility that a wormhole can connect two different universes or two separated spacetime regions in our universe was considered by Morris and Thorne~\cite{Thorne1,Thorne2}, Hawking and Ellis~\cite{Hawking} and Visser~\cite{Visser}. The problem of finding a wormhole solution connecting to the same asymptotic region can at best be done in an axially symmetric solution.

An explicit construction of a charged wormhole which does connect to the same asymptotic region was demonstrated by Schein and Aichelburg~\cite{Schein,Aichelburg}. The wormhole does not violate the positive energy conditions and {\it it does not require negative energy} to stabilize the wormhole throat. In the following, we will adapt their wormhole solution to our modified gravity in which the charge $Q$ is proportional to mass: $Q=\sqrt{\alpha G_N}M$. This means that the interior black hole solution is described by the MOG metric with only the mass $M$ as source, and the exterior repulsive force that stabilizes two connected spheres $S_i\,(i=1,2)$ is due to the gravitational repulsive potential $\phi_\mu=(\phi_0,0,0,0)$. If the stable traversable wormhole only allows one-way travel through it, a time-reversed stable wormhole can be constructed in the same spacetime region that would allow a hapless space adventurer to return to his original destination.

The Majumdar-Papapetrou solution of the MOG field equations (\ref{phiFieldEq}), (\ref{Bequation}) and (\ref{Bcurleq}) allows for a system of bodies to be held in equilibrium by a balance between the gravitational $B_{\mu\nu}$ field repulsion and gravitational attraction~\cite{Majumdar,Papapetrou}. In Cartesian coordinates the metric is given by
\begin{equation}
ds^2=V^{-2}dT^2-V^2(dx^2+dy^2+dz^2),
\end{equation}
where $V(x,y,z)$ satisfies Laplace's equation:
\begin{equation}
\Delta V(x,y,z)=\biggl(\frac{\partial^2}{\partial x^2}+\frac{\partial^2}{\partial y^2}+\frac{\partial^2}{\partial z^2}\biggr)V(x,y,z)=0.
\end{equation}

Two interior non-intersecting spheres $S_i$ are cut out of the Majumdar-Papapetrou spacetime and the potential function $V$ is constant on the surfaces of the spheres. This problem is analogous to determining the electric potential outside two charged metal spheres. The two-body problem is constructed by choosing the z-axis to point along the line of symmetry joining the two spheres $S_i$ with radius $R$. The two spheres are centered at $z=\pm d_1$ and the function $V$ is given by
\begin{equation}
V=V_0=1+\frac{m_1}{R}.
\end{equation}
For large distance separation the two spheres appear as two particles with $Q=M=m_1$ (in relativistic units $G_N=c=1$ and $\alpha(1+\alpha)=4\pi$ or $\alpha=3.08$). The metric potential then has the form~\cite{Schein,Aichelburg}:
\begin{equation}
V(x)=1+\Sigma_{n=1}^\infty\frac{m_n}{|{\bf x}\pm {\bf d}_n|}.
\end{equation}
New coordinates are defined by
\begin{equation}
R=\frac{\tilde r}{\sinh\mu_0},\quad d_1={\tilde r}\coth\mu_0,
\end{equation}
and the polar coordinates:
\begin{equation}
\coth\mu=\frac{(x^2+y^2+z^2+{\tilde r}^2)}{(2{\tilde r}z)},\quad \cot z=\frac{(x^2+y^2+z^2-{\tilde r}^2)}{(2{\tilde r}\sqrt{x^2+y^2})},\quad \cot\phi=\frac{x}{y}.
\end{equation}
The metric now takes the form:
\begin{equation}
ds^2=V^{-2}dT_+^2-V^2\frac{{\tilde r}^2}{(\cosh\mu-\cos\eta)^2}(d\mu^2+d\eta^2+\sin^2\eta d\phi^2),
\end{equation}
and the potential $V$ is given by~\cite{Schein,Aichelburg}:
\begin{equation}
V(\mu,\eta,\phi)=1+\frac{m_1}{R}\biggl[1+\sqrt{\cosh\mu-\cos\eta}\Sigma^{+\infty}_{n=-\infty}(-1)^{n+1}\frac{1}{\sqrt{\cosh(\mu+2n\mu_0)-\cos\eta}}\biggr].
\end{equation}

We can now construct a wormhole by attaching different asymptotic regions of our MOG spacetime to the surfaces $\mu=\pm\mu_0$. This is accomplished by placing two infinitely thin shells of $Q=\sqrt{\alpha G_N}M$ gravitationally charged matter at the transition surfaces of the spheres $S_i$. The interior metric of the shells is given by
\begin{equation}
\label{MOGmetric2}
ds^2=\biggl(1-\frac{2GM}{r}+\frac{\alpha G_NGM^2}{r^2}\biggr)dT^2-\frac{dr^2}{1-2GM/r+\alpha G_NGM^2/r^2}-r^2d\Omega^2.
\end{equation}

We have two asymptotic regions $I$ and $II$ lying outside the MOG spacetimes, and we cut off the asymptotically flat parts. We introduce spherical polar coordinates $(T, r, \theta,\phi)$ centered at $z=-d_1$ so that the radius of the sphere $S^+_1$ is given by $r_+=R$. Then, the metric (\ref{MOGmetric2}) is
\begin{equation}
ds^2=V^{-2}dT_+^2-V^2(dr_+^2+r_+^2d\Omega^2).
\end{equation}
In terms of the proper time $\tau$ the induced metric on the shell $S_1$ is
\begin{equation}
ds^2=d\tau^2-(RV_0)^2d\Omega^2.
\end{equation}
By introducing a second shell $S_2^-$ in the asymptotic region $II$, lying in the causal future of $S_1^-$, a traversable wormhole can be constructed~\cite{Schein,Aichelburg}. Because of the symmetry of the over all system, the shell energy densities and pressures are the same for both shells.

The energy density and pressure obtained for the inner MOG region are given in the limit of $M\rightarrow 0$ by
\begin{equation}
\sigma=-\frac{1}{4\pi V_0^2}\frac{\partial V}{\partial r_+}\Big|_{r_+=R}=-\frac{(\cosh\mu_0-\cos\eta)}{4\pi{\bar r}V_0^2}\frac{\partial V}{\partial\mu}\Big|_{\mu=-\mu_0},\quad p=0.
\end{equation}
The sign of $\partial V/\partial\mu|_{\mu=-\mu_0}$ can be shown to be negative on $S_1$ for arbitrary values of the positive parameter $m_1$~\cite{Schein,Aichelburg}, so that the energy density $\sigma > 0$, thereby, avoiding any violations of positive energy theorems.

\begin{figure}
\centering\includegraphics[scale=0.35]{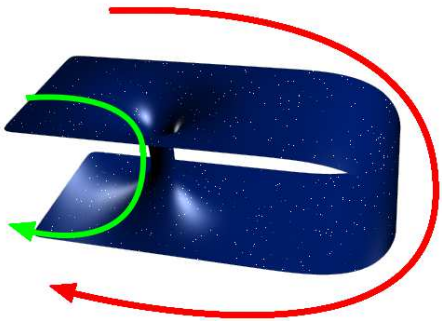}\\
\caption{\label{fig:bridge}A MOG wormhole with a bridge connecting two asymptotically flat spacetimes (from Wikipedia).}
\end{figure}

Any spacelike slice which avoids the singularities in the MOG metric cuts $S_1$ and $S_2$ and connects two separated asymptotic regions. An observer can enter the wormhole through $S_1$ and causally emerge at $S_2$ arbitrarily far in the past or future. In Fig.~\ref{fig:bridge} a MOG wormhole is displayed. A second $S_3$ and $S_4$ construction the same as the one for $S_1$ and $S_2$ in the external Majumdar-Papapetrou region can be constructed some distance away. Then, the second wormhole can be used by the observer to causally get back to his/her original starting place.\\

\section{Conclusions}

A solution is derived for a point particle with mass $M$, gravitational charge $Q=\sqrt{\alpha G_N}M$, and
the gravitational field equations for a spread out distribution of $B_{\mu\nu}$ energy density $T_{\phi\mu\nu}$. As in the case of the Reissner-Nordstr\"om electric charge solution and the Kerr solution, the metric has two horizons. However, if a mechanism such as quantum gravity is invoked to remove the singularity at $r=0$, then this solution could be physically viable as a description of the black hole. A classical MOG solution is derived which is regular at $r=0$ and has two horizons or no horizon.

A Kruskal-Szekeres maximal analytic set of coordinates is determined for both the MOG singular and non-singular (at $r=0$) black hole solutions. The Kerr-MOG solution for a point particle with mass $M$ and angular momentum parameter $a$ can produce a photon sphere which can differ observationally from the photon sphere produced by a pure Kerr metric solution~\cite{Moffat8}. The lensing of the black hole solutions and the images of a black hole shadow can provide observational signatures which can distinguish between the standard GR and MOG black hole solutions. Future observational data obtained by the Event Horizon Telescope (EHT)~\cite{Fish,Lu,Johannsen,Psaltis} can potentially accomplish this task.

A traversable wormhole is constructed by adapting the Schein-Aichelburg solution~\cite{Schein,Aichelburg} to the modified gravitational field equations. The field external to two connected spheres $S_1$ and $S_2$ is described by the Majumdar-Papapetrou solution~\cite{Majumdar,Papapetrou}. The MOG gravitational repulsion balances the attractive gravity resulting in a stable wormhole throat, allowing an observer to traverse from one asymptotically flat region of spacetime to another distant one. This could have the consequence of permitting closed timelike curves and the construction of a time-machine violating causality. However, a quantum description of time travel which avoids closed time-like curves can potentially circumvent a violation of causality~\cite{Seth}.

\section*{Acknowledgments}

I thank Avery Broderick, Martin Green and Viktor Toth for helpful discussions. This research was generously supported by the John Templeton Foundation. Research at the Perimeter Institute for Theoretical Physics is supported by the Government of Canada through industry Canada and by the Province of Ontario through the Ministry of Research and Innovation (MRI).

\end{document}